\documentclass[aps,prl,twocolumn,showpacs,groupedaddress]{revtex4}  % for review and submission 
\usepackage{graphicx}  % needed for figures
\usepackage{dcolumn}   % needed for some tables
\usepackage{bm}        % for math
\usepackage{amssymb,amsmath}   % for math

\begin{document}

% the following line is for submission, including submission to the arXiv!!
%\hspace{5.2in} \mbox{Darmstadt/DESY/U-Tokyo}

\title{Non-thermal fixed points: effective weak-coupling for strongly correlated systems\\ far from equilibrium}
\author{J{\"u}rgen Berges$^{1,2}$}
\author{Alexander Rothkopf$^{3}$}
\author{Jonas Schmidt$^{4}$}
\affiliation{$^{1}$Institute for Nuclear Physics, Darmstadt University of Technology, Schlossgartenstr.\ 9, 64285 Darmstadt, Germany}
\affiliation{$^{2}$Kavli Institute for Theoretical Physics, University of California Santa Barbara, Santa Barbara, CA 93106, USA}
\affiliation{$^{3}$Department of Physics, University of Tokyo, Tokyo 113-0033, Japan}
\affiliation{$^{4}$DESY, Notkestrasse 85, 22603 Hamburg, Germany}

\date{\today}

\begin{abstract}
Strongly correlated systems far from equilibrium can exhibit scaling solutions with a dynamically generated weak coupling. We show this by investigating isolated systems described by relativistic quantum field theories for initial conditions leading to nonequilibrium instabilities, such as parametric resonance or spinodal decomposition. The non-thermal fixed points prevent fast thermalization if classical-statistical fluctuations dominate over quantum fluctuations. We comment on the possible significance of these results for the heating of the early universe after inflation and the question of fast thermalization in heavy-ion collision experiments.  
\end{abstract}

\maketitle

\section{\label{sec:intro} Introduction}

Important phenomena in early universe cosmology ("Big Bang") and
collison experiments of heavy nuclei ("Little Bangs") involve quantum fields far from equilibrium. A prominent topical example concerns the role of nonequilibrium instabilities for the process of thermalization. The heating of the early universe after inflation may proceed via an instability such as parametric resonance~\cite{Traschen:1990sw,Berges:2002cz}. Similarly, plasma instabilities may play an important role in our understanding of observations at the Relativistic Heavy Ion Collider~\cite{Arnold:2004ti}. Instabilities also arise in many other areas, such as dynamics of ultra-cold quantum gases.

Nonequilibrium instabilities lead to exponential growth of field fluctuations on time scales much shorter than the asymptotic thermal equilibration time. Though their origin can be very different, the subsequent evolution after an instability follows similar patterns: After a fast initial period of exponential growth the dynamics slows down
considerably. At this stage all processes become of order unity and one is dealing with a strongly correlated system that has to be treated non-perturbatively, even if the underlying microscopic theory is weakly coupled. The subsequent evolution is characterized by power-law behavior reminiscent of turbulence. It has been argued that
this behavior does not occur in the non-perturbative regime and a perturbative  analysis is employed with apparent success~\cite{Micha:2002ey,Arnold:2005qs}.

In this Letter we show that far-from-equilibrium dynamics in the non-perturbative regime can approach scaling solutions with a dynamically generated weak coupling. 
As an example we consider scalar $N$-component quantum field theory with quartic self-interaction following a parametric resonance instability. In the non-perturbative regime we find new scaling solutions with strongly enhanced low-momentum fluctuations $\sim p^{-4}$ as compared to a high-temperature distribution $\sim p^{-1}$. They correspond to non-thermal fixed points of the time evolution equations for correlation functions once classical-statistical fluctuations dominate over quantum fluctuations. 
At sufficiently high momenta we recover the perturbative behavior $\sim p^{-3/2}$ reported in the literature. 

We employ the two-particle irreducible (2PI) large-$N$ expansion to next-to-leading order~\cite{Berges:2001fi}. The 2PI approximation schemes have been applied to a variety of far from equilibrium phenomena, including parametric resonance~\cite{Berges:2002cz} and tachyonic preheating~\cite{Arrizabalaga:2004iw}. They are known to describe the late-time approach to thermal equilibrium characterized by Bose-Einstein or Fermi-Dirac distributions, respectively~\cite{Berges:2001fi,Berges:2002wr}. The non-perturbative regime after an instability is traditionally described using classical-statistical field theory simulations, and we present a comparison of quantum and classical evolution.

\section{Nonequilibrium instabilities}
\label{sec:inst}

We consider a relativistic real scalar $N$--component quantum field $\Phi_a$ ($a\!=\!1,\ldots, N$) with $\lambda/(4! N)\, (\Phi_a\Phi_a)^2$ interaction, where summation over repeated indices is implied. The macroscopic field of the quantum theory is given by $\phi_a(x) = \langle \Phi_a(x) \rangle$, where the brackets describe the
trace for given initial density matrix. There are two independent two-point correlation functions, which can be associated to the anti-commutator and the
commutator of two fields:
\begin{eqnarray}
F_{ab}(x,y) &=& \frac{1}{2}\langle \left\{ \Phi_a(x), \Phi_b(y)\right\}\rangle -
\phi_a(x) \phi_b(y) \, , 
\nonumber\\
\rho_{ab}(x,y) &=& \langle [\Phi_a(x),\Phi_b(y)]\rangle \, .
\end{eqnarray}
Here $\rho$ is the spectral function, which is related to the retarded propagator
$G_R(x,y) = \rho(x,y) \Theta(x_0-y_0)$. The statistical function $F$ is proportional to "occupation number", which may be taken as $n(t,{\bf p}) + 1/2 = [F(t,t';{\bf p}) \partial_t\partial_{t'} F(t,t';{\bf p})]^{1/2}_{t=t'}$ for spatially homogeneous systems~\cite{Berges:2001fi}.

The nonequilibrium time evolution of $F$ and $\rho$ is described
by coupled differential equations, 
\begin{eqnarray}
\left[\square_x\delta_{ac}+M^2_{ac}(x) \right] \rho_{cb}(x,y)&=&\int_{x_0}^{y_0} \!\!{\rm d}^d z \Sigma_{ac}^\rho(x,z)\rho_{cb}(z,y),
\nonumber\\ 
\left[ \square_x\delta_{ac}+M^2_{ac}(x) \right] F_{cb}(x,y)&=&
\int_{t_I}^{y_0}\!\!{\rm d}^d z \Sigma_{ac}^F(x,z)\rho_{cb}(z,y)
\nonumber\\
&-&\int_{t_I}^{x_0}\!\!{\rm d}^d z \Sigma_{ac}^\rho(x,z)F_{cb}(z,y),
\nonumber\\
 \label{eq:evolution}
\end{eqnarray}
and a similar equation for $\phi_a(x)$~\cite{Berges:2001fi}. These evolution equations would be exact for known effective mass term $M^2(\phi,F)$, spectral (imaginary) part of the self-energy, $\Sigma^{\rho}(\phi,F,\rho)$, and statistical (real) part, $\Sigma^{F}(\phi,F,\rho)$. Here $t_I$ describes the initial time and we will consider a pure initial quantum state with spatially homogeneous fields 
\begin{eqnarray}
\phi_a(t) = \sigma(t) \sqrt{6 N/\lambda}\, \delta_{a1}
\end{eqnarray} 
for $\lambda \ll 1$. All quantities will be given in units of
the initial $\sigma_0 \equiv \sigma(t=0)$. The initial two-point functions
are taken to be diagonal with $F_{ab}= {\rm diag} 
\{F_{\parallel},F_{\perp},\ldots,F_{\perp}\} \sim {\cal O}(N^0 \lambda^0)$ and the initial spectral function is fixed by the equal-time commutation relations. 

Dynamics of parametric resonance in quantum field theory has been studied in detail~\cite{Berges:2002cz} using the 2PI $1/N$ expansion to NLO~\cite{Berges:2001fi}. Primary resonant amplification occurs for the dominant transverse modes in the momentum range 
$0 \le {\bf p}^2 \le \sigma_0^2/2$. This is followed by a non-linear regime, where $F_{\perp} \sim {\cal O} (N^{1/3}\lambda^{-2/3})$. Here occupied low-momentum modes act as sources for a secondary stage of enhanced amplification in a broad higher momentum range. The exponential growth stops when 
$F_{\perp} \sim F_{\parallel} \sim {\cal O} (N^0\, \lambda^{-1})$ and the dynamics slows down considerably. At this non-perturbative stage all processes are of order unity and the system is strongly correlated. 

\begin{figure}[t]
\includegraphics[scale=0.35,angle=-90]{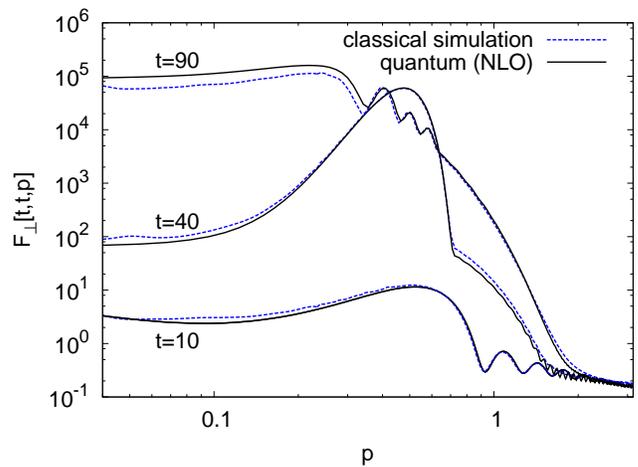}
\caption{\label{fig:comp} Two-point function $F_{\perp}(t,t;{\bf p})$ as a function of momentum $|{\bf p}|$ for three different times. Quantum (solid) and classical (dashed) evolution agree well even where $F_{\perp}({\bf p}) \ll 1$.}
\end{figure}
Fig.~\ref{fig:comp} shows $F_{\perp}(t,t,{\bf p})$ for times $t=10$, $t=40$ in the non-linear regime, and $t=90$ in the non-perturbative regime. The solid line shows the result for the quantum evolution for $\lambda = 0.01$ and $N=4$. For comparison, the dashed line gives the same quantity as obtained from simulations of the corresponding classical-statistical field theory on a lattice with same initial conditions following Ref.~\cite{Aarts:2001yn}. The precision of agreement between the curves is remarkable. Quantum fluctuations are expected to be suppressed if the classicality condition $F^2 \gg \rho^2$ is fulfilled~\cite{Aarts:2001yn}. However, with $\rho^2$ being of order unity the quantum corrections turn out to be extremely small even for $F^2 \ll 1$. We emphasize that this is not a generic property of the NLO approximation~\cite{Berges:2001fi,Aarts:2001yn}, but a consequence of the instability dynamics and $\lambda \ll 1$. In turn, contributions beyond NLO seem to play an inferior role even for the non-perturbative regime.

\section{Non-thermal fixed points}

Fixed points are time and space translation invariant solutions of the evolution equations (\ref{eq:evolution}), which require
\begin{eqnarray}
\Sigma^{\rho}_{ac}(p) F_{cb}(p) - \Sigma^{F}_{ac}(p) \rho_{cb}(p) &\equiv& \left(J_{ab}^{(3)}(p) + J_{ab}^{(4)}(p)\right)/\lambda
\nonumber\\
&\stackrel{!}{=}& 0 \, .
\label{eq:fp}
\end{eqnarray}
Thermal equilibrium solves (\ref{eq:fp}) using the fluctuation-dissipation relation $F^{\rm (eq)}_{ab}(p) = \left[ n_{T}(p_0) + 1/2 \right] \rho^{\rm(eq)}_{ab}(p)$ and  correspondingly for self-energies~\cite{Berges:2001fi}. We show (\ref{eq:fp}) has non-thermal approximate solutions if classical-statistical fluctuations dominate over quantum fluctuations.    

\begin{figure}[t]
\vspace*{3.ex}
\includegraphics[scale=0.3]{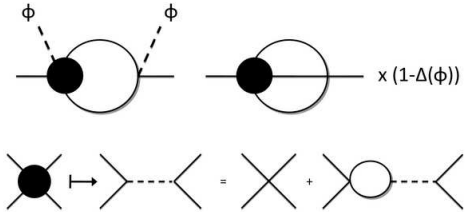}
\caption{\label{fig:graphs} NLO contributions to the fixed point equation (\ref{eq:fp}).}
\end{figure}

\begin{figure}[t]
\includegraphics[scale=0.35,angle=-90]{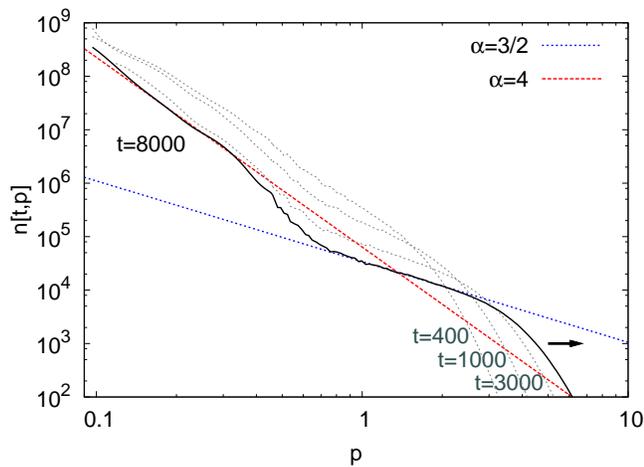}
\caption{\label{fig:occup} Occupation number as a function of momentum for different times. The solid line corresponds to $t=8000$, along with fits $\sim |{\bf p}|^{-4}$ for low and $\sim |{\bf p}|^{-3/2}$ for higher momenta.}
\end{figure}

Firstly, we analytically determine fixed point solutions with
$F_{ab}(p) = f(p)/\lambda\, \delta_{ab}$, $\rho_{ab}(p) = \rho(p) \delta_{ab}$ and
$\sigma = {\rm const} \neq 0$. Secondly, these non-trivial solutions are shown to describe well the slow dynamics in the non-perturbative regime. At NLO of the 2PI $1/N$ expansion in the classical-statistical field theory limit we have 
\begin{eqnarray}
J_{aa}^{(3)}(p;\sigma) &=& - \frac{\sigma^2}{3 (2 \pi)^4}
\int\! {\rm d}^4 k\, {\rm d}^4 q\, \delta^{4}(p-q-k)
\nonumber\\
&& \left[\lambda_{\rm eff}(k) + \lambda_{\rm eff}(q) + \lambda_{\rm eff}(p)\right] [\rho(k) f(q) f(p)
\nonumber\\
&&  + f(k) \rho(q) f(p) - f(k) f(q) \rho(p)] \, ,
\label{eq:j3nlo}
\end{eqnarray}
where we summed over components $a$. Here
\begin{eqnarray}
\lambda_{\rm eff}(p) = \frac{1}{\left|1+\Pi_R(p)\right|^2}
\label{eq:leff}
\end{eqnarray}
with the resummed "one-loop" retarded self-energy
\begin{eqnarray}
\Pi_R(p) = \frac{1}{3 (2 \pi)^4} \int\! {\rm d}^4 q\,  f(q) G_R(p-q) \, .
\label{eq:rs}
\end{eqnarray}
Diagrammatically, $J^{(3)}$ contains contributions described by the "one-loop" graph in Fig.~\ref{fig:graphs}. In contrast to perturbative Feynman diagrams, here lines correspond to self-consistently dressed propagators $F$ or $\rho$~\cite{Berges:2001fi}. The full dot denotes the effective vertex (\ref{eq:leff}), which is graphically displayed in the lower graph of Fig.~\ref{fig:graphs}. It iteratively generates the infinite series of resummed "chain" diagrams contributing at NLO in the $1/N$ expansion.
Similarly,
\begin{eqnarray}
J_{aa}^{(4)}(p;\sigma) &=& \frac{1}{18 (2 \pi)^8}
\int\! {\rm d}^4 k\, {\rm d}^4 q\,
{\rm d}^4 r\, \delta^{4}(p+k-q-r)
\nonumber\\
&& \lambda_{\rm eff}(p+k) \Big\{[f(p) \rho(k) + \rho(p) f(k)]
f(q) f(r)
\nonumber\\
&&  - f(p) f(k) [f(q) \rho(r) + \rho(q) f(r)] \Big\}
\nonumber\\
&& \left( 1 - \Delta(p+k;\sigma) \right)
\label{eq:j4nlo}
\end{eqnarray}
corresponds to the "two-loop" graph in Fig.~\ref{fig:occup} with
$\Delta(p;\sigma) = 4 \sigma^2 {\rm Re}[G_R(p)/(1+\Pi_R(p))]$.

\begin{figure}[t]
\includegraphics[scale=0.35,angle=-90]{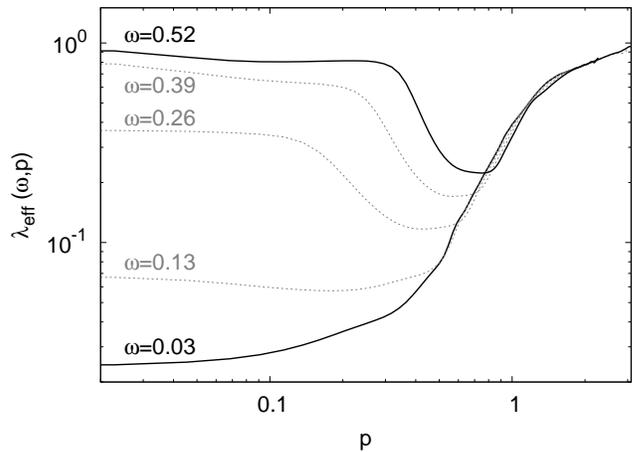}
\caption{\label{fig:leff} The effective vertex at $t=240$ as a function of spatial momentum ${\bf p}$ for different frequencies $\omega \equiv p_0$.}
\end{figure}

In the non-perturbative regime, where the dynamics is slow, one can extract 
quantities such as $\lambda_{\rm eff}(p)$ also directly from the nonequilibrium quantum evolution by Fourier transformation with respect to relative coordinates. The result is shown in Fig.~\ref{fig:leff} for $t=240$ as a function of three-momentum $\bf p$ for different values of the frequency $\omega \equiv p_0$, with parameters as in Fig.~\ref{fig:comp}. For momenta larger than order unity $\lambda_{\rm eff}(p)$ tends to one. One observes that this holds if either $\bf p$ or $\omega$ is not small. Therefore, in this range the dynamics is well approximated by the one- and two-loop expressions as displayed in Fig.~\ref{fig:graphs} without the effective vertex. Most strikingly, for small four-momentum the infinite series of ${\cal O}(1)$ contributions add up to a substantially reduced effective coupling. The latter is still slowly changing in time and is shown to approach a power-law behavior in the following.

For scaling solutions the correlators behave under simultaneous scaling $p \to sp$ and $\phi \to s\phi$ as~\footnote{Symmetries allow a non-trivial dynamical exponent $z$ with
$F(p_0,{\bf p}) = s^{2+\alpha}\, F(s^z p_0, s {\bf p})$ etc. Our solutions are consistent with $z=1$ and we employ this. Keeping $z$ explicity the fixed point solutions (\ref{eq:irexp}) below read $\alpha = 3z + 1$ and $\alpha = 4z + 1$ which reduce to the given values for $z=1$.}:
\begin{eqnarray}
F(p) &=& s^{2+\alpha}\, F(sp) \, ,\nonumber\\
\rho(p) &=& s^2\, \rho(sp) \, ,
\label{eq:scaling}\\
\lambda_{\rm eff}(p) &=& s^{\gamma}\, \lambda_{\rm eff}(sp) \, .
\nonumber
\end{eqnarray}
Similarly, we have $G_R(p) = s^2 G_R(sp)$ and it follows that $\Pi_R(p) = s^\alpha \Pi_R(sp)$. 
Taking $s p \sim 1$ one concludes for $\alpha > 0$ that $\Pi_R(p) \gg 1$ for $p \to 0$. According to (\ref{eq:leff}) the infrared scaling behavior of the effective vertex in this case is described by (\ref{eq:scaling}) with the exponent
\begin{eqnarray}
 \gamma = - 2 \alpha \, .
\label{eq:irgamma}
\end{eqnarray}
Using this scaling behavior we find   
$J_{aa}^{(3)}(p;\sigma) = s^0\, J_{aa}^{(3)}(s p;s \sigma)$ and
$J_{aa}^{(4)}(p;0) = s^{\alpha}\, J_{aa}^{(4)}(s p;0)$
such that $J_{aa}^{(4)}(p;0)$ dominates the infrared dynamics. The field-dependent part $J_{aa}^{(4)}(p;\sigma) - J_{aa}^{(4)}(p;0)$ scales as $J_{aa}^{(3)}(p;\sigma)$ since $\Delta(p;\sigma) = s^{-\alpha} \Delta(s p;s \sigma)$ in this regime. 

Accordingly, for the infrared we consider solutions of 
$\int {\rm d}^3 p J_{aa}^{(4)}(p_0,{\bf p};0) = 0$.
The integration over spatial momentum allows us to solve this equation using generalized Zakharov transformations~\cite{Zakharov}. In this way the problem is reduced to simple algebraic conditions for the exponents. E.g., to map the second term in the integrand sum of (\ref{eq:j4nlo}) onto the first we employ the transformation $k_0 \to p_0^2/k_0$,
$q_0 \to q_0 p_0/k_0$, $r_0 \to r_0 p_0/k_0$ as well as ${\bf p} \to {\bf k} p_0/k_0$, ${\bf k} \to {\bf p} p_0/k_0$, ${\bf q} \to {\bf q} p_0/k_0$, ${\bf r} \to {\bf r} p_0/k_0$, and similar for the third and fourth term.
Using the scaling properties (\ref{eq:scaling}) the integrand vanishes if $\alpha = (4-\gamma)/3$ or $\alpha = (5-\gamma)/3$. With (\ref{eq:irgamma}) this yields two non-thermal fixed point solutions for the infrared scaling behavior:
\begin{eqnarray}
\alpha = 4 \,\, , \,\, \alpha = 5 \, .
\label{eq:irexp}
\end{eqnarray}
We do not list classical thermal solutions ($\alpha=1$, $0$), which appear as a consequence of the fluctuation-dissipation relation discussed above.
We emphasize that (\ref{eq:irexp}) cannot be obtained from a perturbative $2 \leftrightarrow 2$ scattering analysis~\cite{Zakharov,Micha:2002ey}, which we recover for vanishing $\gamma$. 
 
The size of the one-loop retarded self-energy (\ref{eq:rs}) distinguishes the low-momentum from a high-momentum regime. The latter is defined by $\Pi_R(p) \ll 1$ and in this range the effective vertex scales trivially ($\gamma = 0$) according to (\ref{eq:leff}) with $\lambda_{\rm eff}(p) \simeq 1$. Therefore, at high momenta 
$J_{aa}^{(3)}(p;\sigma) = s^{2 \alpha}\, J_{aa}^{(3)}(s p;s \sigma)$ and 
$J_{aa}^{(4)}(p;0) = s^{3 \alpha}\, J_{aa}^{(4)}(s p;0)$
such that $J_{aa}^{(3)}(p;\sigma)$ dominates over $J_{aa}^{(4)}(p;0)$ for $\alpha > 0$.
The field-dependent part $J_{aa}^{(4)}(p;\sigma) - J_{aa}^{(4)}(p;0)$ scales in the same way as $J_{aa}^{(4)}(p;0)$ since here $\Delta(p;\sigma) = s^0 \Delta(s p;s \sigma)$. A similar analysis as above yields for $\int {\rm d}^3 p J_{aa}^{(3)}(\omega,{\bf p};\phi) = 0$ in this regime
\begin{eqnarray}
\alpha &=& 3/2 \, .
\label{eq:uvexp}
\end{eqnarray}
Consequently, simulation results may be fitted by perturbative estimates down to rather low momenta~\cite{Micha:2002ey}.

It remains to show that the nonequilibrium evolution approaches non-thermal fixed points. For this we follow the classical-statistical evolution numerically to late times. Fig.~\ref{fig:occup} gives the occupation number $n(t,{\bf p})$ as a function of three-momentum at different times. We employ $N=4$ and a weak coupling $\lambda = 10^{-5}$ in order to keep quantum corrections small. The latest time shown is represented by the solid line. For a region of momenta above $|{\bf p}| \gtrsim 1$ the evolution is very well approximated by a power-law behavior $n(t,{\bf p})\sim |{\bf p}|^{-3/2}$ in agreement with (\ref{eq:uvexp}). As time proceeds this region grows towards higher momenta, which is indicated by the arrow in Fig.~\ref{fig:occup}. 

Going to smaller momenta one first observes a transition region as expected from the above discussion. The infrared behavior is consistent with a power-law that is well approximated by $\alpha = 4$. The evolution in this region is very slow, as can be inferred from comparing to the $t=3000$ line, and getting an even better possible agreement between numerics and analytics would be very costly in computational time. 
From the current data the $\alpha = 5$ solution indicated in (\ref{eq:irexp}) is clearly unfavored. 

Scaling solutions govern the nonequilibrium dynamics only if classical-statistical fluctuations dominate~\footnote{Other approximate fixed points may also be present~\cite{Aarts:2000wi}.}. For initial conditions leading to instabilities the statistical fluctuations grow large. In this case the system approaches non-thermal fixed points exhibiting critical slowing down. Non-thermal fixed points are unstable with respect to quantum corrections, which eventually lead to thermalization. The quantum evolution is diagrammatically described by same topologies as displayed in Fig.~\ref{fig:graphs} at NLO. However, pairs of propagators $f(p) f(q)$ in classical equations (\ref{eq:j3nlo}) or (\ref{eq:j4nlo}) can be associated to $f(p) f(q) + (\lambda^2/4) \rho(p) \rho(q)$ in the quantum theory~\cite{Berges:2001fi}.  The coupling does not drop out and no universal scaling solutions appear if quantum corrections dominate.  

Non-thermal fixed points can excessively delay thermalization. In the context of early universe reheating a most conservative limit requires thermal equilibrium at a temperature of order $10$ MeV before Big Bang Nucleosynthesis. Even this might already rule out some very weakly coupled inflaton models~\cite{Micha:2002ey}, however, more  realistic models have to be considered including quantum corrections. In the context of heavy-ion collisions, QCD plasma instabilities can lead to a regime with qualitatively similar scaling behavior~\cite{Arnold:2005qs}. Classical simulations indicate possible fast isotropization due to instabilities in about $1$-$2$ fm/c for low momenta of less than $1$-$2$ GeV~\cite{Berges:2007re}. However, subsequent scaling behavior leads to large isotropization times for higher momentum modes in this case, which seems incompatible with experimental findings. So far, the role of quantum corrections for the lower occupied high-momentum modes were not taken into account. The question of whether it is possible to find also an effective weak coupling in QCD, which may facilitate an analytic quantum description, is exciting.

We are indebted to Szabolcs Bors{\'a}nyi for many discussions and computational help.

\end{document}